\documentclass{article}
\usepackage{fleqn,ams}
\usepackage{amsmath,amssymb}
\usepackage{epsfig}
\usepackage{verbatim}
\newcommand{\const}{\mbox{Const}}

\begin{document}

\begin{center}
{\bf\large The moving semibounded magnetoactive plasma\\[2pt] in
field of a flat gravitational wave}\\[12pt]
Yu.G. Ignatyev, E.G. Chepkunova\\
Kazan State Pedagogical University,\\ Mezhlauk str., 1, Kazan
420021, Russia
\end{center}

\begin{abstract}
The problem of a moving semibounded magnetoactive plasmas in a
plane gravitational wave  is research on the basis of the
self-consistent equations, obtained earlier by the
Authors.\end{abstract}

\section{Introduction}
In the series of articles (\cite{gmsw}, \cite{gmsw2},
\cite{gmsw3}, \cite{gmsw4}), based on exact stationary1 solutions
of the set of Maxwell equations and the equations of relativistic
magne\-to\-hyd\-ro\-dy\-na\-mics for an unlimited homogeneous
plasma, we established an essentially nonlinear behaviour of a
highly magnetized plasma in the field of gravitational radiation
propagating across the magnetic field. On reaching some critical
con\-di\-ti\-ons, the plasma begins accelerating up to subluminal
velocities in the direction of the gravitational wave propagation.
Simultaneously, the magnetic field strength greatly increases.
This article is devoted to an analysis of the boundary-value
problem for a semibounded plasma, with a goal to study the
mechanism of the origin of a gravimagnetic shock wave.

\section{Self-consistent equations of motion
of a magneto active plasma in the PGW field}
In Ref. \cite{gmsw}, from the condition of coincidence of dynamic
velocities in the energy-momentum tensors (EMT) of a perfect fluid
and the electromagnetic field, the equations of relativistic
magnetohydrodynamics of plasma in the arbitrary gravitational
field were obtained, and their exact solution was found in the
background of the plane gravitational wave (PGW) metric,
corresponding to an initial state of a homogeneous electroneutral
plasma with a frozen-in homogeneous magnetic field.

A general property of the above solution is the presence of a
singularity on the hypersurface
\begin{equation}
\label{ii4} \Delta(u)=1-\alpha^{2}\left[e^{2\beta(u)}-1\right]=0,
\end{equation}
ãäå
\begin{equation} \label{ii5}
\alpha^{2}=H_{0}^{2}\sin^{2}\Omega/ 4\pi(\varepsilon_{0}+p_{0})
\end{equation}
is a dimensionless parameter while $\beta(u)$ is an arbitrary
function of the retarded time u (the PGW amplitude). This
singularity in a weak PGW ($\beta \ll 1$) is reached under the
condition:
\begin{equation}
\label{ii6} 2\beta(u)\alpha^{2} \geq 1\, .
\end{equation}
An analysis of the solution shows that, when the condition
(\ref{ii6}) holds, a shock wave is formed in the plasma,
propagating with a subluminal velocity in the direction of PGW
propagation. A necessary condition for the excitation of a shock
magnetohydrodynamic wave (GMSW) is a high magnetization of the
plasma:
\begin{equation}
\label{ii7} \alpha^{2} \gg 1\ .
\end{equation}
However, the above solution of the RMHD equations is essentially
stationary (it depends only on u, the retarded time) and
corresponds to an initially homogeneous magnetoactive plasma.
Therefore the solution obtained in Ref. \cite{gmsw3}] cannot
describe the dynamics of the shock wave excitation mechanism.

In this paper, we will study the problem of PGW distribution in an
isotropic magnetoactive plasma with boundary conditions, supposing
that the null hyper\-sur\-face:
\begin{equation} \label{sigma0}
\Sigma_0 : u = 0
\end{equation}
is the surface of PGW front, i.e., the PGW is absent at $u \leq
0$, --
\begin{equation}
\label{ii2.23} \beta(u)_{|u\leq 0} = 0\,;\hspace{0.5cm}
\beta'(u)_{|u\leq 0} = 0\,; \hspace{0,5cm} L(u)_{|u\leq 0} = 1\,.
\end{equation}
In the absence of a PGW, the plasma is at rest, i.e.,
\begin{equation}
\label{ii2.25a} \psi(u,v)_{|u\leq 0} = \psi_{0}(x^{1})\,;
\end{equation}
\begin{equation}
\label{ii2.25} p_{|u\leq 0} = p_{0}(x^{1})\,; \hspace{1cm}
\varepsilon_{|u\leq 0} = \varepsilon_{0}(x^{1})\,,
\end{equation}
where $\psi_{0}\,, p_{0}\,, \varepsilon_{0}$ - are some given
functions of the variable $x^{1}$;
\begin{equation}
\label{ii2.26} v_{v |u\leq 0} = v_{v|u\leq 0} = \frac{1}{\sqrt{2}}
\Longrightarrow -
\left(\frac{\partial_{u}\psi}{\partial_{v}\psi}\right)_{|u\leq 0}
= 1\,.
\end{equation}

Thus, due to (\ref{ii2.26}):
\begin{equation}
\label{ii2.27} - (\partial_{u}\psi)_{|u\leq 0} =
(\partial_{v}\psi)_{|u\leq 0}\,.
\end{equation}

In the absence of a PGW, the energy density, pressure and magnetic
field strength are identical everywhere:
\begin{equation}
\label{iig1} \varepsilon_{|u\leq 0} = \varepsilon_{0} =
\mbox{Const} \,;\hspace{1cm} p_{|u\leq 0} = p_0 = \mbox{Const}\,;
\end{equation}
\begin{equation} \label{iig2}
H_{|u\leq 0} = H_{0} = \mbox{Const} \Rightarrow \psi_{|u\leq 0} =
\frac{1}{\sqrt2}H_0 (v - u)\\,
\end{equation}
at $u\leq 0\,; \forall\ v\ \in {\cal R}$.

In Ref. [4] it has been shown that the plasma macroscopic
parameters are the following:
\begin{equation}
\label{ii3.6} v_{u} = \sqrt{ - \frac{\partial_{u}Z}{
\partial_{v}Z}}\,; \hspace{0.5cm} v_{v} = \sqrt{ - \frac{\partial_{v}Z}{
\partial_{u}Z}}\,.
\end{equation}
\begin{equation}
\label{ii3.11} H^{2} = -L^{-4}e^{2\beta}\partial_u Z \partial_v Z
\,,
\end{equation}
\begin{equation} \label{ii3.15}
\varepsilon = L^{-2}\varepsilon_0\sqrt{-\partial_u Z
\partial_v Z}
\end{equation}
where
\begin{equation} \label{iig50}
Z = \frac{\sqrt 2\psi}{H_0}
\end{equation}
is a dimensionless function.

The function $Z(u,v)$ should satisfy a boundary condition on the
spacelike hypersurface  $\Sigma_r : u = v \,(x=0).$) Let us put,
following Ref. \cite{gmsw}, on this hypersurface:
\begin{equation}\label{iig2a}
H_{|u=v}={\rm Const}. \end{equation}
Then, taking into account the relation (see Ref.  \cite{gmsw}):
\begin{equation} \label{ii2.6} H_{2} =
F_{13} = \frac{1}{\sqrt{2}} (F_{v3} - F_{u3}) = \frac{1}{\sqrt{2}}
(\partial_{v} \psi - \partial_{u} \psi )\,.
\end{equation}
we bring the boundary condition (\ref{iig2a}) to the form:
\begin{equation} \label{iig10}
- \left(\partial_{u}Z \partial_{v}Z\right)_{|u=v} = 1\,.
\end{equation}

Taking into account these conditions, the set of RMHD equations is
reduced to a single differential equation:
$$ \partial_{uv}Z + \beta'\partial_{v}Z  = \frac{2
\pi e^{-2 \beta}}{H^2_0}\frac{dp}{d\varepsilon}(\varepsilon + p)
\frac{(L^{2})'}{\partial_{u}Z}- $$
$$ -\frac{\pi L^2}{H^2_0} e^{-2
\beta}(\varepsilon + p)
\Biggl\{\left[\frac{\partial_{vv}Z}{(\partial_{v}Z)^{2}} +
\frac{\partial_{uu}Z} {(\partial_{u}Z)^{2}} -
2\frac{\partial_{uv}Z}{\partial_{u}Z
\partial_{v}Z}\right]+$$
\begin{equation}\label{iig6}
+\frac{dp}{d\varepsilon}
\left[\frac{\partial_{vv}Z}{(\partial_{v}Z)^{2}} +
\frac{\partial_{uu}Z} {(\partial_{u}Z)^{2}} +
2\frac{\partial_{uv}Z}{\partial_{u}Z
\partial_{v}Z}\right]\Biggl\}
\end{equation}
which for a non-relativistic plasma becomes
\begin{eqnarray} \label{g7} \partial_{uv}Z +
\beta'\partial_{v}Z = - \frac{1}{4\alpha^2} e^{- 2\beta} \sqrt{-
\partial_{v}Z
\partial_{u}Z}\times \nonumber\\
\times\left[\frac{\partial_{vv}Z}{(\partial_{v}Z)^{2}} +
\frac{\partial_{uu}Z} {(\partial_{u}Z)^{2}} -
2\frac{\partial_{uv}Z}{\partial_{u}Z
\partial_{v}Z}\right]\,.
\end{eqnarray}
(for details see Ref. \cite{gmsw4}).

\section{Search for a solution in a weak
gravitational wave}
Since the GMSW is formed in a highly magnetized plasma, in what
follows we shall suppose that $\alpha$ (\ref{ii5}) is a large
parameter (see the condition (\ref{ii6})). Therefore, Eq.
(\ref{g7}), being expanded in the small dimensionless parameter of
the problem $\alpha^{-1}$, takes the form:
\begin{equation}
\label{g7a} \partial_{uv}Z + \beta'\partial_{v}Z =0
\end{equation}
Solving it, we obtain the first approximation, satisfying the
conditions indicated above:
\begin{equation} \label{no3}
Z_0 = (v - u) \, e^{-\beta}- 2{\displaystyle \int _{0}^{u} \sinh
(\beta (u))\,du}
\end{equation}
The magnetic field behaviour corresponding to the solution
(\ref{no3}) is shown in Fig.\ref{h}:.
\vskip 6pt\noindent
\begin{center}
\parbox{8cm}{\refstepcounter{figure} \label{h}
\epsfig{file=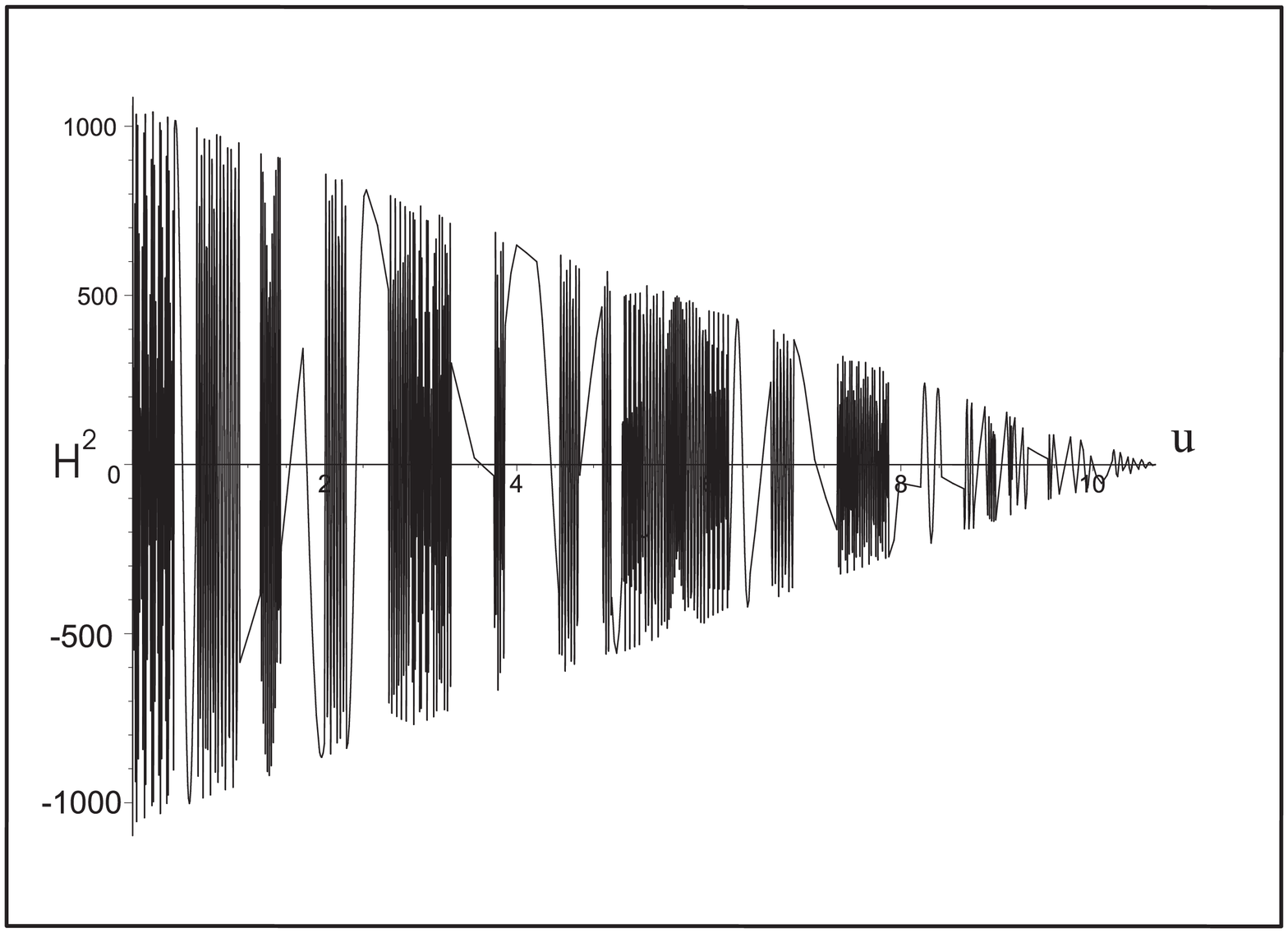, angle=0,width=8cm } \vskip 0.5cm \noindent
Figure \thefigure: \hskip 6pt {\sl Dependence of the magnetic
field intensity on the retarded time calculated according to the
solution (\ref{no3}).\hfill} }\end{center}

To verify the linear approximation applicability, we substitute
the solution found to the r.h.s. of Eq. (\ref{g7}) and obtain:
\begin{equation}
\label{g7b} \partial_{uv}Z + \beta'\partial_{v}Z = \Phi(u,v),
\end{equation}
where $\Phi(u,v)$ is a very unwieldy expression. It is, however,
easy to see that the radicand $-\partial_u{Z_0}\partial_v{Z_0}$
can be negative on some surface $\Sigma(u,v)$ (an analysis shows
that it is $\partial_u Z_0$ that changes its sign). It means that
near this surface the linear solution of Eq. (\ref{no3}) becomes
inapplicable. Supposing that the GW amplitude ƒÀ is everywhere
small,
\begin{equation}
\label{beta} |\beta(u)| \ll 1
\end{equation}
and decomposing the radicand at the r.h.s. of Eq. (\ref{g7b}) in
series with respect to $\beta$, we obtain the equation of this
surface:
\begin{equation}
\Sigma(u,v): \quad (v-u)\beta'+\beta+1=0,
\end{equation}
which shows that this surface exists at sufficiently large values
of the variable $(v-u)= \sqrt{2} x$, i.e., it is far from the
boundary:
\begin{equation}
|v-u|\sim \frac{1}{\beta_0\omega},
\end{equation}
where $\beta_0$ is the GW amplitude and $\omega$ its frequency.

Near the surfaces $\Sigma(u,v)$:
\begin{equation}
1+\beta' (v-u)=\sigma \ll 1,
\end{equation}
whence, putting $\beta(u)=\beta_0(1-\cos \omega u)$, we obtain the
equation of the surface $\Sigma(u,v)$, resolved explicitly with
respect to retarded time:
\begin{equation}
\label{v} v=\frac{-1+\sigma}{\beta_0 \omega \sin(\omega u)+u};
\end{equation}

The ranges in which Eq. (\ref{v}) holds are represented by narrow
parabolas which have almost flat vertices lying near the straight
line $u=v+a \,(a=\const)$. Their qualitative behaviour is depicted
in Fig. \ref{v(u)_a}:

\vskip 6pt\noindent
\begin{center}
\parbox{8cm}{\refstepcounter{figure} \label{v(u)_a}
\epsfig{file=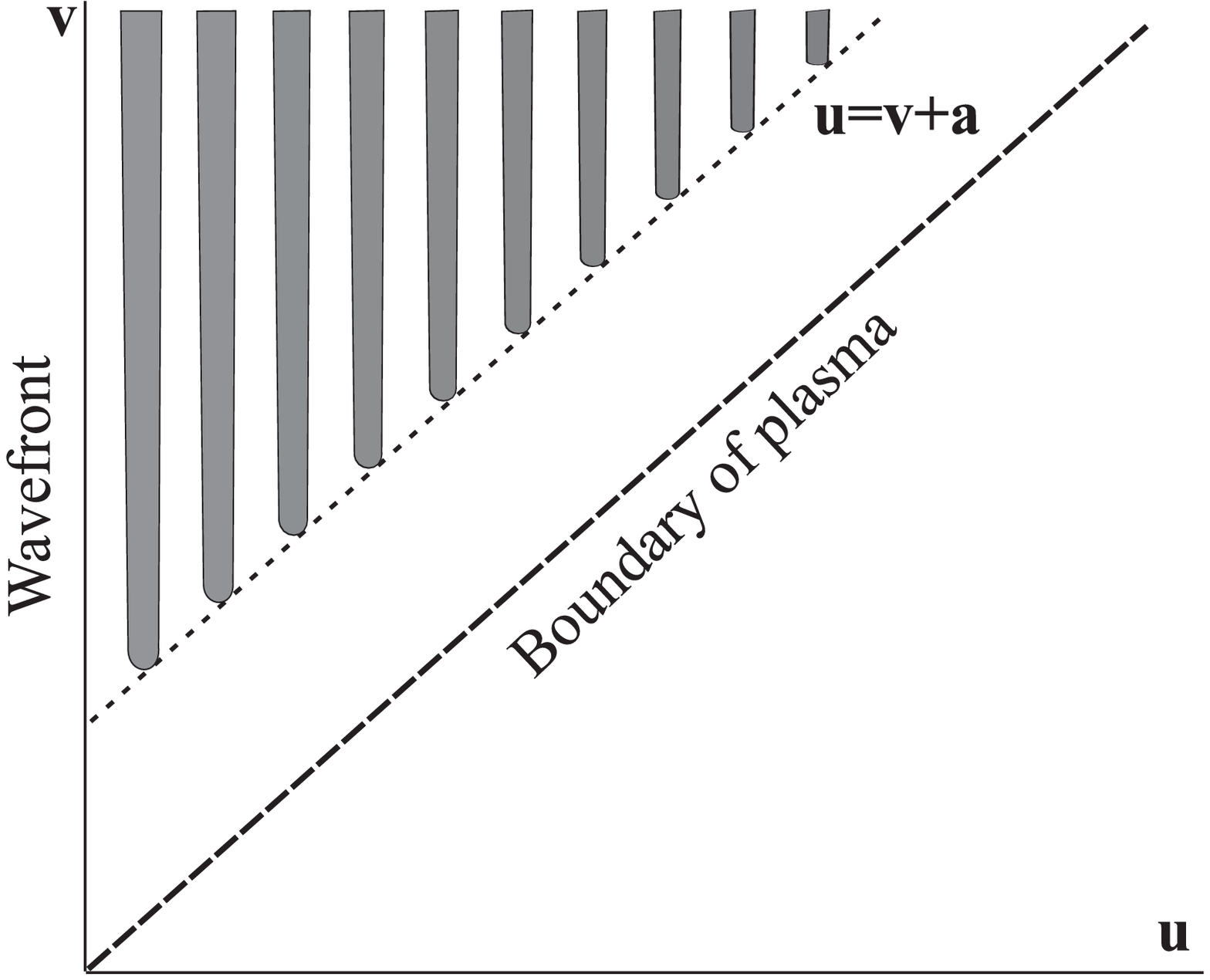, angle=0,width=6cm } \vskip 0.5cm \noindent
Figure \thefigure: \hskip 6pt {\sl Ranges of violation of the
linear approximation of Eq. (\ref{g7a})\hfill} }
\end{center}

A detailed image of the vertex of one of the parabolas is shown in
Fig. \ref{v(u)}.
\vskip 6pt\noindent
\begin{center}
\parbox{8cm}{\refstepcounter{figure} \label{v(u)}
\epsfig{file=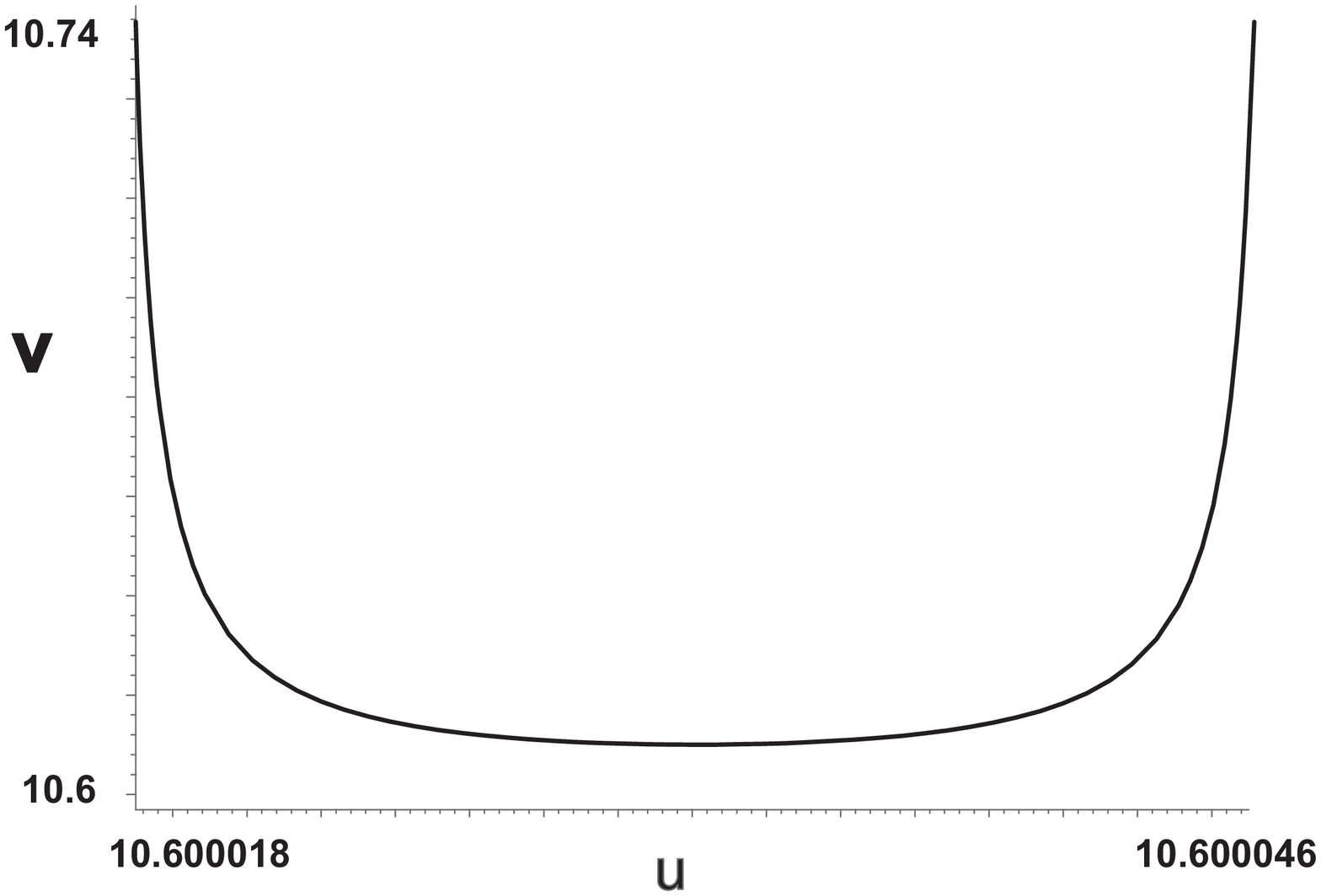, angle=0,width=6cm } \vskip 0.5cm \noindent
Figure \thefigure: \hskip 6pt {\sl The dependence v(u) inside one
of ranges, shown in Fig. \ref{v(u)_a}.\hfill} }\end{center}

Far from this surface, Eq. (\ref{g7b}) is integrated, and we
obtain the first-order correction:
\begin{equation}
\label{no4} Z_1 = (v - u) \, e^{-\beta} \zeta + 2{\displaystyle
\int _{0}^{u} \zeta \cosh (\beta (u))\,du}
\end{equation}
where
\begin{equation}
\label{zeta}\zeta=\int_{0}^{u} \Phi(u) e^{\beta(u)}\,du.
\end{equation}

Since an analytical solution to Eq. (\ref{g7}) cannot be found,
while a direct application of numerical methods faces considerable
difficulties, related to large values of the derivatives near
singular points, we have used, for a numerical solution of Eq.
(\ref{g7}), the symmetric reflection method, approximating the
solution behaviour near a singular point with a symmetric
parabola.

Comparing the resulting numerical solution of Eq. (\ref{g7}) near
singular points with the analytical solution (\ref{no3}) of Eq.
(\ref{g7a}), we conclude that, despite the difficulties indicated
above, these solutions almost coincide. An illustration is given
in Fig. \ref{z}:
\vskip 6pt\noindent
\begin{center}
\parbox{8cm}{\refstepcounter{figure} \label{z}
\epsfig{file=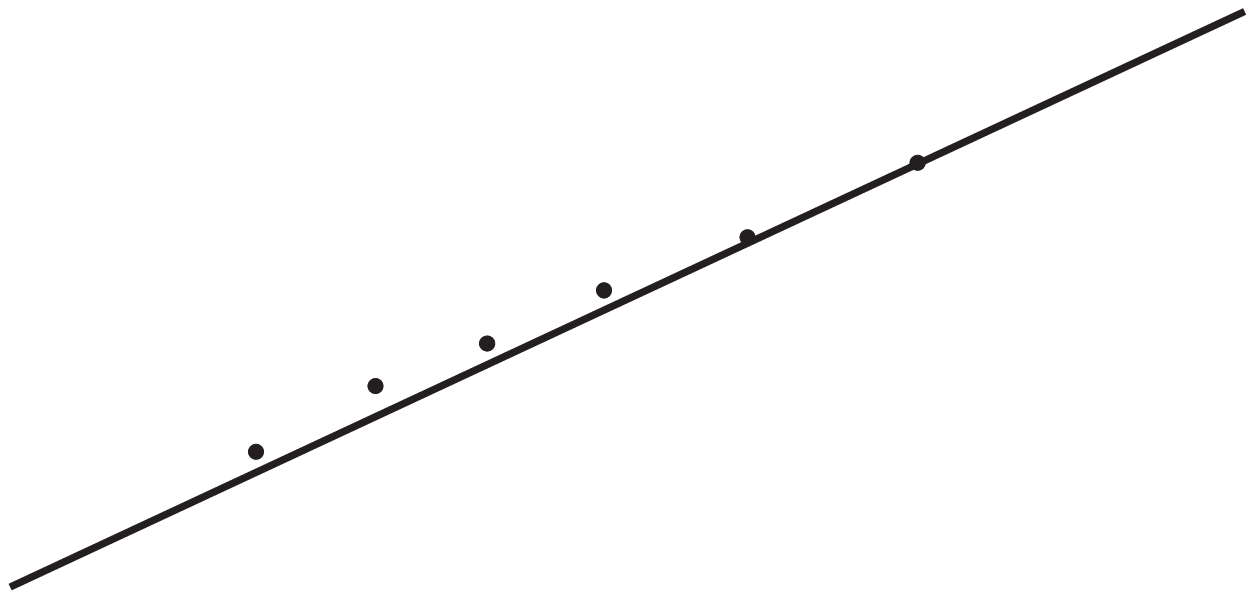, angle=0,width=8cm } \vskip 0.5cm \noindent
Figure \thefigure: \hskip 6pt {\sl Comparison of the numerical and
analytical solutions at small scale inside a range shown in
\ref{v(u)_a}: the straight line represents an analytical solution,
the points give a numerical solution. \hfill} }
\end{center}
\ref{h^2} shows a qualitative pattern of the magnetic field
strength squared near a singular point. As a result, taking into
account the nonlinearity of Eq. (\ref{g7}) near singular2 points
reduces, basically, to a cut-off of the lower range ($H^2<0$) in
the graph \ref{h} and to formation of a plateau of the function
$H^2$ near zero. Let us remark that a similar behaviour of
magnetoactive plasma inside a nonlinear range was also established
in Ref. \cite{21}.
\vskip 6pt\noindent
\begin{center}
\parbox{8cm}{\refstepcounter{figure} \label{h^2}
\epsfig{file=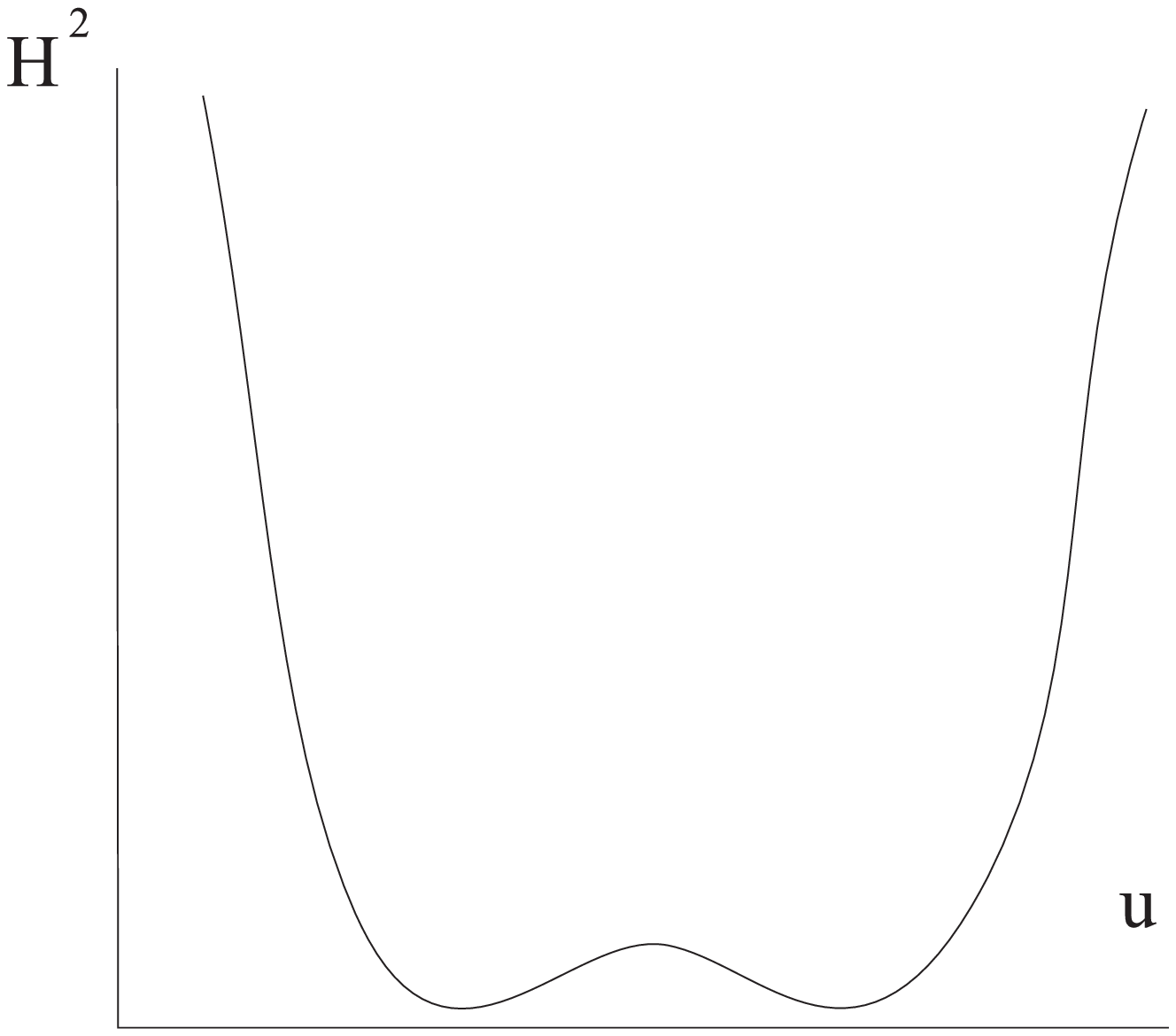, angle=0,width=6cm } \vskip 0.5cm
\noindent{\bf Ðèñ. \thefigure.} \hskip 6pt {\sl Magneticfield
behaviour near singular points\hfill} }
\end{center}

As a result, taking into account the nonlinearity of Eq.
(\ref{g7}) near singular2 points reduces, basically, to a cut-off
of the lower range ($H^2<0$) in the graph \ref{h} and to formation
of a plateau of the function $H^2$ near zero. Let us remark that a
similar behaviour of magnetoactive plasma inside a nonlinear range
was also established in Ref. \cite{21}.

Let us also remark that, in a weak gravitational wave ($\beta \ll
1$), the condition $\partial_v Z_0 \approx 1$ is satisfied, and
therefore Eq. (\ref{g7}) can be reduced to the form of a
first-order quasi-linear partial differential equation:
\begin{equation}
\label{zv_1} \partial_u \varphi * \varphi^{-3/2} + 4 \alpha^2
\partial_v \varphi = \beta'
\end{equation}
where $\varphi=-\partial_u Z$.

It is easy to obtain the formal common solution of this equation,
however, since it is impossible to find this solution explicitly,
it is also impossible to find the function $Z(u,v)$ satisfying the
required initial and boundary conditions.

\end{document}